\documentclass[12pt,dvips]{article}
\usepackage[dvips]{graphicx}
\usepackage{graphicx}
\usepackage{epsfig}
\usepackage{amsmath,amssymb,amsfonts}
\usepackage{cite}
\usepackage{axodraw4j}
\usepackage{subfigure}
\usepackage[x11names]{xcolor}

\graphicspath{{plots/}}
\DeclareGraphicsExtensions{.eps.gz,.eps,.ps,.ps.gz}

\parskip=\itemsep               %?
\newcommand{\lambdabar}{{\hbox{$\lambda_e$\kern-1.9ex\raise+0.45ex\hbox{--}
\kern+0.2ex}}}
\newcommand{\mg}{m_{\gamma^{\prime}}}

\def\beq{\begin{equation}}
\def\eeq{\end{equation}}

\newif\ifhepph

\ifhepph\date{\empty}\fi

\title{
{
\normalsize
\rightline{
CERN-BE-2009-035; DESY 09-121} \rightline{DCPT/09/114; IPPP/09/57}}\
\vskip 0.cm
\LARGE{\textbf{Feasibility, engineering aspects and physics  reach of microwave cavity
experiments searching for hidden
photons and axions}} \\
\vspace{11mm}
}

\author{
Fritz Caspers$^{1,}$\footnote{{\bf e-mail}: fritz.caspers@cern.ch}\,\,,
Joerg Jaeckel$^{2,}$\footnote{{\bf e-mail}: joerg.jaeckel@durham.ac.uk}\,\,,
 and
Andreas Ringwald$^{3,}$\footnote{{\bf e-mail}: andreas.ringwald@desy.de}
\\[2ex]
\small{\em $^1$CERN, CH-1211 Geneva, Switzerland}\\
\small{\em $^2$Institute for Particle Physics and Phenomenology, Durham University, Durham DH1 3LE, UK}\\
\small{\em $^3$Deutsches Elektronen-Synchrotron, Notkestra\ss e 85, D-22607 Hamburg, Germany}
\vspace{1cm}}

\date{}

\begin{document}
\begin{titlepage}
  \maketitle
\begin{abstract}
Using microwave cavities one can build a resonant
``light-shining-through-walls'' experiment to search for hidden sector photons and axion like particles, predicted
in many extensions of the standard model. In this note we make a feasibility study of the sensitivities which
can be reached using state of the art technology.
\end{abstract}

\thispagestyle{empty}
\end{titlepage}
\newpage \setcounter{page}{2}

\section{Introduction}
\begin{figure}[t]
\begin{center}
  \scalebox{0.8}[0.8]{
  \begin{picture}(386,130) (159,-207)
    \SetWidth{1.0}
    \SetColor{Black}
    \GBox(336,-206)(368,-78){0.882}
    \Line[dash,dashsize=8.4,arrow,arrowpos=0.5,arrowlength=5,arrowwidth=2,arrowinset=0.2](272,-142)(432,-142)
    \Photon(160,-142)(272,-142){7.5}{6}
    \Photon(432,-142)(544,-142){7.5}{6}
    \Vertex(432,-142){8}
    \Vertex(432,-142){8}
    \Vertex(272,-142){8}
    \SetOffset(0,-10)
    \Text(224,-110)[c]{\Large{\Black{$\gamma$}}}
    \Text(480,-110)[c]{\Large{\Black{$\gamma$}}}
    \Text(304,-110)[c]{\Large{\Black{$X$}}}
    \Text(400,-110)[c]{\Large{\Black{$X$}}}
  \end{picture}
  }
  \end{center}
\caption{Schematic of a ``light-shining-through a
wall'' experiment. An incoming photon $\gamma$ is converted into a new particle $X$ which interacts only very weakly with
the opaque wall. It passes through the wall and is subsequently reconverted into an ordinary photon which can be detected.
In the case of an axion the conversion is facilitated by a magnetic field that interacts with the incoming photon.
In contrast, photon $\leftrightarrow$ hidden photon oscillations occur via a non-diagonal mass term and can therefore also occur in vacuum (analog to neutrino oscillations).
}\label{Fig:lsw}
\end{figure}
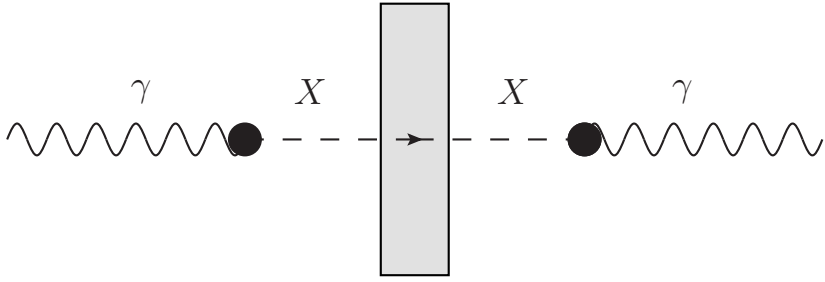

Many extensions of the standard model contain extra light particles with very feeble interactions with the standard model particles -- so called WISPs (for very Weakly Interacing Sub-eV Particles).
Two prominent examples for WISPs are axion like particles (ALPs)~\cite{Peccei:1977hh,Svrcek:2006yi} and hidden sector photons
(HSPs)~\cite{Okun:1982xi,Holdom:1985ag,Dienes:1996zr} (see e.g. Refs.~\cite{Jaeckel:2008ja,Ringwald:2008cu} for reviews).
Apart from astrophysical and cosmological observations (see e.g. Ref.~\cite{Raffelt:1996}), the best laboratory probes exploit the enormous
precision of low energy photon experiments (see e.g. Ref.~\cite{Gies:2007ua,Redondo:2008zf}).
A particularly sensitive technique is a so-called light-shining-through-walls
setup~\cite{Okun:1982xi,Anselm:1986gz,Anselm:1987vj,VanBibber:1987rq}, as shown in Fig.~\ref{Fig:lsw}. In such an experiment
photons are converted into one of those particle species, the latter traverse the wall and are re-converted into photons on the opposite
side of the wall. The wall prevents the unconverted photons from reaching the detector.
In such experiments, the production/regeneration probability can be increased by reflecting the light many times back and forth\cite{Hoogeveen:1990vq,Sikivie:2007qm}.
This corresponds to a high quality
cavity. Cavities with particularly high quality factors can be build in the microwave
regime and promise to be quite sensitive probes of WISPs~\cite{Hoogeveen:1992uk,Jaeckel:2007ch}.
In this note we want to study the technical feasibility and the physics reach of these experiments. In particular, we discuss state of the art microwave detection methods and investigate
to what extend one can probe so-far untested parameter space.
This is a timely enterprise in view of the fact that a number
of microwave cavity experiments are currently set up \cite{Yale} or planned \cite{Daresbury,tobar} at various laboratories around the world.

\section{Setting the stage}

Let us start by looking at the simplest case of a hidden sector photon (HSP) with Lagrangian,
\begin{equation}
\label{masskinmix}
{\mathcal{L}}= -\frac{1}{4}
F^{\mu\nu}F_{\mu\nu}-\frac{1}{4}B^{\mu\nu}B_{\mu\nu}
-\frac{1}{2}\chi\,F^{\mu\nu}B_{\mu\nu} +\frac{1}{2} m^2_{\gamma^{\prime}}
B_\mu B^\mu,
\end{equation}
where $F^{\mu\nu}$ is the ordinary electromagnetic gauge field strength, $B^{\mu}$ is the hidden sector gauge field and $B^{\mu\nu}$ the corresponding gauge field strength.
The HSP mass is $m_{\gamma^{\prime}}$. Finally $\chi$ is the kinetic mixing parameter~\cite{Holdom:1985ag} which induces photon -- hidden photon oscillations
which are analogous to neutrino oscillations (for current constraints, see Fig.~\ref{Fig:hpsensitivity}).

%%%%%%%%%%%%%%%%%%%%%%%%%%%%%%%%%%%%%%%%%%%%%%%%%%%%%%%%
%%%%%%%%%%%%%%%%%%%%%%%%%%%%%%%%%%%%%%%%%%%%%%%%%%%%%%%%

In Ref.~\cite{Jaeckel:2007ch} it was shown that, in case of HSP induced microwave-shining-through-walls,
the power output ${\mathcal{P}}_{\rm{det}}$ of the receiver cavity (with quality factor $Q^\prime$) is related to the
power input ${\mathcal{P}}_{\rm{em}}$ of the emitter cavity (with quality factor $Q$) by the
expression
\begin{equation}
\label{ptrans}
{\mathcal{P}}_{\rm{det}}
= \chi^4\,  \frac{\mg^{8}}{\omega^{8}_{0}}\,  |G|^2\,Q Q^{\prime}\,{\mathcal{P}}_{\rm{em}},
\end{equation}
where $\omega_0=2\pi f_0$ is the resonance frequency
of the cavity. The geometrical details of the setup are encoded in the factor
\begin{equation}
\label{geofac}
G(k/\omega_{0})\equiv \omega^{2}_{0}\int_{V^{\prime}}\int_{V}d^{3}\mathbf{x}\,d^{3}\mathbf{y}\,
\frac{\exp({\rm i} k |\mathbf{x}-\mathbf{y}|)}{4\pi|\mathbf{x}-\mathbf{y}|}
A_{\omega_{0}}(\mathbf{y})A^{\prime}_{\omega_{0}}(\mathbf{x}),
\end{equation}
where $V$ and $V^{\prime}$ are the volumina of the emitter and receiver cavities and $A_{\omega_{0}}$ and $A^{\prime}_{\omega_{0}}$
are the normalized eigenmodes of the cavities.
A very similar expression holds for ALPs,
\begin{equation}
\label{ptransalp}
{\mathcal{P}}_{\rm{det}}
\sim \left( \frac{g\, B}{\omega_0}\right)^4\,  |\tilde{G}|^2\,  Q Q^{\prime}\,{\mathcal{P}}_{\rm{em}},
\end{equation}
where $g$ is the coupling of ALPS to two photons\footnote{For a scalar or pseudoscalar ALP this coupling corresponds to a term
$-\frac{g}{4}\phi_{+} F^{\mu\nu}F_{\mu\nu}$ or $-\frac{g}{4}\phi_{-} F^{\mu\nu}\tilde{F}_{\mu\nu}$, respectively.} and $B$ denotes the magnitude of an external
magnetic field which is needed in this case since, unlike HSPs, ALPs have spin-0, prohibiting
photon-ALP oscillations in vacuum. $\tilde{G}$ is a geometry factor similar to Eq.~\eqref{geofac}.

In any case, the measurable quantity from which we want to extract the signal is the power output ${\mathcal{P}}_{\rm{det}}$ of the receiver cavity. To optimize the sensitivity of
the experiment we can now try to
\begin{itemize}
\item[a)]{increase this power output and/or}
\item[b)]{improve the sensitivity of the detector in order
to detect smaller power outputs from the receiver cavity.}
\end{itemize}
Effectively, a) means optimizing the right hand side of the expressions~\eqref{ptrans} or
\eqref{ptransalp}. Here, we can try to increase the $Q$s, the emitter
power or try to optimize $|G|$.

The latter can be done only to a limited degree. Geometry factors much bigger than $1$ are difficult to achieve. But in realistic setups with not too large
distances between emitter and receiver cavities and using low modes of the cavities values between $|G|\sim 10^{-3}$ and $1$ are doable.

One can increase the quality of the emitter cavity $Q$, but what effectively counts is the total amount of energy stored inside the cavity which is
proportional to $Q{\mathcal{P}}_{\rm{em}}$. Typically cavities have a lower $Q$ when operated at high field strength, so one should directly optimize the
relevant quantity which is the energy stored inside. For example, the cavities originally designed for the TESLA accelerator~\cite{Aune:2000gb,Lilje:2004ib}
achieve  $Q{\mathcal{P}}_{\rm{em}}\sim10^{10}\times 10\,{\rm W}$.

Finally, we can try to optimize the quality $Q^{\prime}$ of the receiver cavity. In principle, this quality factor is the quality factor at low field strengths
which can be somewhat higher than the one at high field strength. However,
we have to assure
that both cavities have the same resonant frequency. More precisely
the frequencies have to agree in a small range
$\Delta\omega_{0}/\omega_{0}\sim 1/Q^{\prime}$. This is a
non-trivial task. However, compared to optical frequencies (as
proposed in \cite{Hoogeveen:1990vq,Sikivie:2007qm}), this should be
significantly simpler for microwave or RF cavities: the wavelength
is longer and correspondingly the allowed inaccuracies in the cavity
are much larger. Indeed, the cavities originally developed for the
TESLA accelerator~\cite{Lilje:2004ib} may be mutually tuned in
frequencies to a ${\rm{few}}\times 100\,\rm{Hz}$~\cite{Knabbe:priv}.
With a resonance frequency of roughly 1 GHz, this corresponds to an
allowed quality factor of the detector cavity of $Q^{\prime}\sim
10^{6}$. Such $Q$ values are also reachable with normal conducting cavities,
the latter having the advantage also to be usable with an external magnetic
field which is mandatory for the ALP search.

This leaves us with optimizing the detector to which we turn in the next section.

\section{Narrowband microwave detection methods}

One advantage of the microwave-shining-through-walls setup is that the frequency of the signal must be exactly the same frequency as used to produce the new particles.
This can be used to suppress noise -- and therefore improve signal to noise -- by using a
narrowband detection method. We will illustrate this method by sketching the the principle of a lock-in amplifier (for a brief review see, e.g.,~\cite{lockin}).
Later on we will describe a more comfortable and more powerful relative of this method using a Fast Fourier Transform (FFT) signal analyzer
(for an introduction see, e.g.,~\cite{FFT,agilent}),.

A lock-in amplifier takes the input signal -- in our case the power coupled out of the receiver cavity --
and mixes it with (i.e. multiplies it by) the reference signal -- in our case taken from the generator of the
field in the emitter cavity. In fact, a true signal caused by WISPs would follow exactly the frequency of the generator. Therefore, a suitable reference signal can be obtained by taking a small part of the generator power.
The lock-in amplifier then integrates the mixed signal over time.
Because sine and cosine functions of different frequencies are orthogonal when integrated over infinitely long times this procedure gives a non-vanishing signal only if
the input frequency matches the reference frequency. For finite integration times $\Delta t$, the matching has to be within a bandwidth given by $BW=1/\Delta t$.
For example a sine reference signal and an input waveform $U_{\rm in}(t)$, the DC output signal $U_{\rm out}(t)$ can be calculated for an analog lock-in amplifier by:
\begin{equation}
U_{\mathrm{out}}(t)= \frac{1}{\Delta t} \int_{t-\Delta t}^t {\sin\left[\omega_{\mathrm{ref}}\cdot t^{\prime} + \phi\right] U_{\mathrm{in}}(t^{\prime})}\;\mathrm{d}t^{\prime}
\end{equation}
where $\omega_{\rm ref}$ is reference frequency and $\phi$ is a phase that can be set on the lock-in. Using an input signal $U_{\rm in}(t)=U_{\rm in} \sin(\omega^{\prime}t)$
we have
\begin{eqnarray}
U_{\rm out}(t)\!\!&=&\!\!\frac{1}{2\Delta t(\omega_{\mathrm{ref}}-\omega^{\prime})}[\sin((\omega_{\mathrm{ref}}-\omega^{\prime})t+\phi)-\sin((\omega_{\mathrm{ref}}-\omega^{\prime})(t-\Delta t)+\phi)]
\\\nonumber
&&\!\!\!\!\!\!+\frac{1}{2\Delta t(\omega_{\mathrm{ref}}+\omega^{\prime})}[\sin((\omega_{\mathrm{ref}}+\omega^{\prime})t+\phi)-\sin((\omega_{\mathrm{ref}}+\omega^{\prime})(t-\Delta t)+\phi)].
\end{eqnarray}
This function is strongly peaked around $\omega^{\prime}=\pm\omega_{\mathrm{ref}}$. The width of the peak is $\sim 1/\Delta t$.
Therefore the lock-in amplifier acts as a bandpass of width
\begin{equation}
\label{width}
BW=1/\Delta t.
\end{equation}

To obtain the signal to noise ratio we have to compare this output signal
to the one caused by white noise in the receiver,
\begin{equation}
\label{SNR}
{\rm SNR}=\frac{{\mathcal{P}}_{\rm{det}}}{k T_{R} BW}=\frac{{\mathcal{P}}_{\rm{det}} \Delta t}{k T_{R}},
\end{equation}
where $T_{R}$ is the so-called noise temperature of the receiver.
Here, we note that the SNR grows linearly with the measurement time $\Delta t$.
This advantageous property arises because with the lock-in technique the effective
bandwidth affected by the white noise
decreases with time according to \eqref{width}.
We stress that this improvement relies on our knowledge of the phase of the produced hidden photons and in turn the regenerated ordinary photons
in the detector. This is in contrast to the situation, for example, in axion dark matter experiments~\cite{Sikivie:1983ip} where the phase and precise frequency of the
searched for axions is in principle unknown.

So far we have considered a single measurement with a measurement time $\Delta t$.
Since the variance of the noise is of the size of the noise a ${\rm SNR}=1$ corresponds to a $1\sigma$ signal.
To improve the SNR one can in principle average over
a number of measurements $N$. This reduces the \emph{variance} of the noise by a factor of
$\sqrt{N}$ (the average noise power, of course, remains the same).
And accordingly the significance improves by a factor $\sqrt{N}$. Averaging a number of times we can therefore clearly detect
even signals with ${\rm SNR}\lesssim 1$.

Let us start with a modest setup at room temperature $300$~K.
At this temperature the white noise corresponds to
$k T\sim 4\times 10^{-21}$~W/Hz$=-174$~dBm/Hz.
After a measurement time of 1000~s we have a bandwidth of $10^{-3}$~Hz. Therefore, the total power
of noise within the bandwidth is $k T BW=4\times10^{-24}$~W$=-204$~dBm.
Therefore in an idealized setup we have a sensitivity of
\begin{equation}
{\mathcal{P}}^{{\rm SNR}=1}_{\rm{det}}=4\times 10^{-24}\, {\rm W},\quad\quad\quad{\rm at\,\,300\,K}.
\end{equation}
This corresponds to less than 2 photons per second at 5 GHz.
At the end of this section we will give an example of a setup where a sensitivity of $10^{-22}$~W has been achieved in practise with a measurement
time of $300$~s.
In this context it should be noted that, of course, we do not really detect single photons but we profit from the huge statistics which
can be accumulated over a long measurement period.
Let us stress here, that this is an idealized calculation and that in a realistic situation one probably has to measure longer and/or average over a number of
measurements.

In a more advanced setup at cryogenic temperatures we can assume noise temperatures of the order of
$10$~K. Combining this with integration times of the order of
$1000$~s we theoretically achieve a SNR$=1$ (in one measurement) with a signal power,
\begin{equation}
{\mathcal{P}}^{{\rm SNR}=1}_{\rm{det}}=1.4\times 10^{-25}\, {\rm W},\quad\quad\quad{\rm at\,\,10\,K}.
\end{equation}
At a frequency of 5 GHz this corresponds to a tiny flux of $0.04$ photons per second.

At the moment it seems that we could improve the sensitivity by measuring for longer and longer times. One might
wonder whether there is a principal limitation to the achievable sensitivity. Indeed there is. As we have noted above
the lock-in method relies on the knowledge of the phase (or more precisely the phase difference between sender and detector)
of the signal to detect. However, because the mass of the hidden photons (or ALP) is a priori unknown this leads to an uncertainty
in the phase of the detectable particle. The time delay between a photon traveling the distance $d$ from generator
to detector and a hidden photon is,
\begin{equation}
\delta t=d\left(\frac{1}{v}-1\right)=d\,\frac{1-\sqrt{1-m^2/\omega^2}}{\sqrt{1-m^2/\omega^2}}.
\end{equation}
The uncertainty now arises from the fact that the generator frequency drifts in the time $\delta t$.
If the generator frequency drifts at a rate $\mu_{\rm drift}$ the drift in frequency in the time $\delta t$ and therefore the minimal bandwidth is
\begin{equation}
\label{drift}
BW_{\rm min}\sim \delta \omega=\mu_{\rm drift}\delta t
\sim 0.17 \times 10^{-8}{\rm Hz}\left(\frac{d}{m}\right)\left(\frac{\mu_{\rm drift}}{\rm Hz/s}\right)\left(\frac{m}{\omega}\right)^2.
\end{equation}
In a typical situation the generator drifts very slowly within a 1 Hz window.
For a frequency drift rate $\mu_{\rm drift}=1\,{\rm Hz/s}$ Eq.~\eqref{drift}
corresponds to a measuring time $t_{\rm max}\sim 10^8\, {\rm s}\sim
3$~years. This is quite a bit beyond the currently achievable
measurement times of the order of days.

We are now ready to move towards a more advanced setup involving a Fast Fourier Transform (FFT) device.
As in the lock-in amplifier above we mix, i.e. multiply the signal with the reference frequency.
However, instead of integrating over time one records the full signal and does a Fourier analysis (or more precisely an FFT) of the recorded
signal.

%%%%%%%%%%%%%%%%%%%%%%%%%%%%%%%%%%%%%%%%%%%%%%%%%%%%%%%%
\begin{figure}[t]
\centerline{\includegraphics[width=10cm]{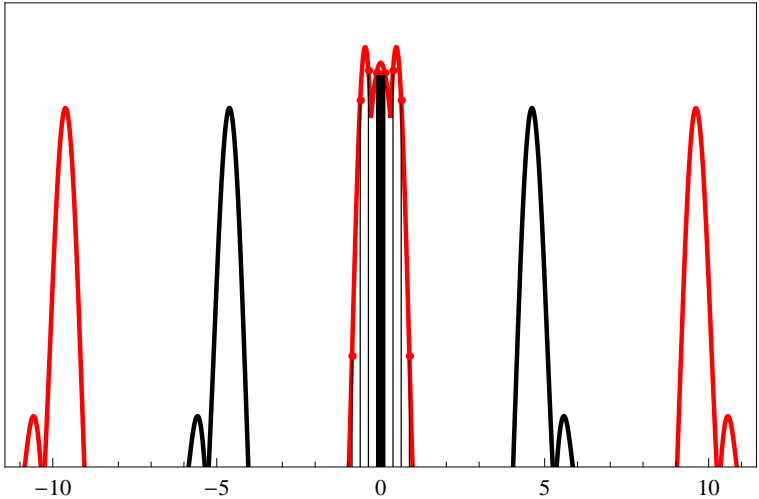} }
\caption{The black line shows the Fourier transform of a signal which is concentrated around the reference frequency $\omega_0=\pm 5$ (arbitrary units).
After multiplying it with the reference frequency the signal is now concentrated around $0,\pm 2\omega_{0}$ (red curves).
Using the a measurement time $\Delta t$ we can then analyze the spectrum with
a resolution $\delta \omega=2\pi/\Delta t$ (thin black lines denote the border of the bins).
The black shaded bins shows the bin measured by a simple lock-in amplifier.}\label{Fig:spectrum}
\end{figure}
%%%%%%%%%%%%%%%%%%%%%%%%%%%%%%%%%%%%%%%%%%%%%%%%%%%%%%%%

To understand the reason for this procedure let us take a look at Figure~\ref{Fig:spectrum}. In Fourier space our signal will typically look
as the black curve. It is centered around the very high reference frequency $\omega_{0}$. In our case this is the frequency of the cavity.
On top of this there are relatively slow time variations some of which we will artificially introduce by modulation. These fall into a small
frequency band around the reference frequency (width of the black curve).
After multiplication with the reference signal the Fourier spectrum will be shifted by $\pm \omega_{0}$. This is shown in the red curve.
Using the FFT we can now analyze the part of the spectrum near $0$ frequency. Due to its high frequency the part centered around $2\omega_{0}$ will average away
and not disturb our measurements (we can also remove it with a bandpass filter).

Simply integrating over the recorded signal would return us to the case of the lock-in amplifier discussed above.
Effectively we would measure the signal strength in the black shaded bin in Fig.~\ref{Fig:spectrum}.
However, performing an FFT we can now analyze the full spectrum with the same resolution $\Delta \omega=2\pi/\Delta t$, i.e.
we obtain the information about all the bins indicated in the figure.
Effectively we are performing a large number of lock-in amplifier measurements with slightly different frequencies simultaneously.
As we will see in the next section this will be very useful to distinguish signal and background.
The noise/sensitivity considerations for the lock-in amplifier now hold for each individual bin.

There is one final ingredient to our FFT. So far we have not discussed the sampling rate of our recorded data. Of course, this rate cannot be infinite.
This will limit the range of frequencies we can analyze. If we sample with a frequency $\omega_{\rm sample}$ we can at best obtain information
in a frequency band of width $\Delta\omega_{\rm FFT}=\omega_{\rm sample}/2$. This follows from the Nyquist-Shannon theorem \cite{nyquist} but becomes immediately plausible when
comparing the number of recorded data points $N_{\rm points}=\omega_{\rm sample}\Delta t/(2\pi)$ to the number of Fourier coefficients $N$ in a frequency band $\Delta\omega_{\rm FFT}$
with frequency resolution $\delta \omega=2\pi/\Delta t$, $N_{\rm coeff}=2\Delta\omega_{\rm FFT}/\delta\omega=2\Delta \omega_{\rm FFT}\Delta t/(2\pi)$.
Effectively this finite bandwidth was also the reason why we first moved the signal to a low frequency by multiplying it with the reference frequency.

\begin{figure}[t]
\begin{center}
%\subfigure{
\begin{picture}(360,150)(0,-40)
\includegraphics[width=12cm]{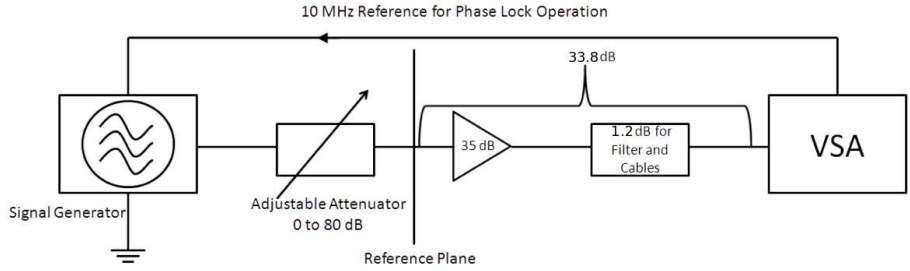}
\end{picture}
%}
%\hspace*{0.8cm}
%\subfigure{
\includegraphics[width=12cm]{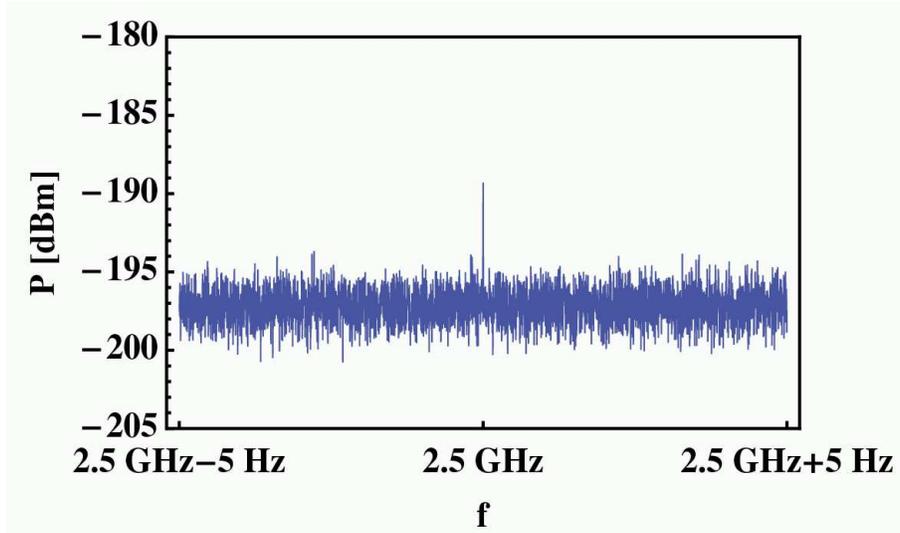}
%}
\end{center}
\caption{In the upper panel a test setup for microwave detection with the methods described in the text (cf. also~\cite{technical}). A weak signal is generated
in the signal generator and is then further attenuated down to $-190$~dBm$=10^{-22}$~W. This signal is then amplified by a total of 33.8 dB and combined (i.e. mixed)
with the reference signal in the vector spectrum analyzer which also records the signal and performs the FFT.
In the lower panel the observed signal averaged over 10 measurements
(this smoothes out the fluctuations of the \emph{envelope} of the narrowband filtered and peak detected noise) is shown (the resolution bandwidth is 3 mHz).
A narrow signal line is observed~\cite{technical} which is clearly
distinct from thermal background (-199 dBm) plus the noise from the amplifier (2dB).}\label{Fig:testsetup}
\end{figure}

Using the setup shown in Fig.~\ref{Fig:testsetup} with a commercially available vector spectrum analyzer\footnote{For the FFT.} (Agilent 9020 MXA)  and a
standard low-noise amplifier
a sensitivity for a detection
of $10^{-22}\,{\rm W}$ with 10 measurements of $300\,{\rm s}$ each has been demonstrated at room temperature~\cite{technical}.
As one can see from the right hand side of Fig.~\ref{Fig:testsetup} the signal sticks out quite clearly from the background and one would expect that
even a 3 (roughly 5 dB) times smaller signal would have been clearly detected.

%%%%%%%%%%%%%%%%%%%%%%%%%%%%%%%%%%%%%%%%%%%%%%%%%%%%%%%%

\section{Distinguishing between a signal and a leak -- Shielding and leakage monitoring}

A crucial ingredient for the experiment is to achieve sufficient shielding.
For the intended sensitivity one has to achieve roughly a shielding of 300 dB (a factor $10^{-30}$) between the emitter  and receiver cavity.
In order to provide sound data, testing of the shielding is mandatory.
Since we  must be absolutely sure that we are not fooled by simple electromagnetic leakage.
Obviously in order to reduce the noise it would also be desirable to have a very good shielding of the emitter section.
Of course, as long as no signal is observed this demonstrates that the shielding works.
However, here we want to be slightly more ambitious and show that one can actually monitor the shielding even during the measurement.

\begin{figure}[t]
\centerline{\includegraphics[angle=-90,width=15cm]{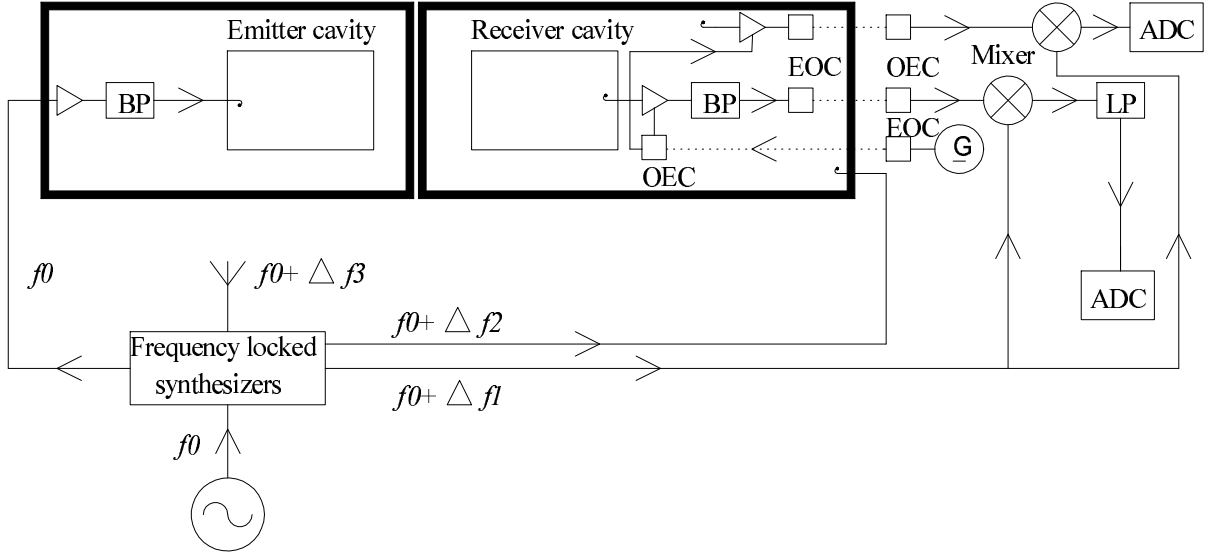} }
\vspace{2.3cm}
\caption{Setup of the experiment with ``test signals'' at slightly different
frequencies for leakage monitoring. For details see main text.}\label{Fig:box1}
\end{figure}

Our proposed setup is shown in Fig.~\ref{Fig:box1}. A suitable way to achieve the required amount of shielding is a box-in-the-box setup.
In the simplest setup the cavity itself provides the internal box and the another layer of external shielding is placed
around it (thick lines in Fig.~\ref{Fig:box1}). If we use cooled cavities and separate cryostats for the emitter and the receiver
the outer layer of shielding is actually already provided by the cryostat.
If more shielding is desired one can add more layers.

%%%%%%%%%%%%%%%%%%%%%%%%%%%%%%%%%%%%%%%%%%%%%%%%%%%%%%%%
%%%%%%%%%%%%%%%%%%%%%%%%%%%%%%%%%%%%%%%%%%%%%%%%%%%%%%%%

Let us now look at the setup in more detail. The generator (bottom of the figure) generates a signal of frequency $f_{0}=\omega_{0}/(2\pi)$.
This signal is then modulated with a set of frequency locked synthesizers\footnote{In a more advanced setup this could also be
done by using serrodyne type frequency shifters which provide a better frequency lock.},
which return, in addition to the signal at the original frequency three modulated
signals with frequencies $f_{0}+\Delta f_{i}$, $i=1,\ldots 3$ (the three other signals are for heterodyne detection and leakage monitoring we will return to this below).
The signal with frequency $f_{0}$ is then fed into the emitter box on the left hand side. Inside the emitter box the signal is amplified and runs through a bandpass filter.
After that it is fed into the cavity. As usual in cavities we can monitor the excited mode and that we are on resonance by checking the reflected wave.

On the receiver (right hand side box) side we proceed similarly. The signal is coupled out of the cavity amplified and passes through a bandpass. However, in
order to minimize the number of cables going in and out of the cavity (to avoid leakage) we then use an electro optical converter (EOC) to change the signal into an
optical signal which is then transmitted to the outside of the box with a glass fibre (dotted lines) where it is reconverted with an opto electrical converter (OEC).
In a similar way the power to the amplifier is fed in with an EOC-OEC set.

\begin{figure}[t]
\begin{center}
\begin{picture}(220,140)
\Text(120,-10)[c]{\scalebox{0.7}[0.7]{$\Delta f_{1}$}}
\Text(160,-10)[c]{\scalebox{0.7}[0.7]{$\Delta f_{1}\!+\!\Delta f_{2}$}}
\Text(210,-10)[c]{\scalebox{0.7}[0.7]{$\Delta f_{1}\!+\!\Delta f_{3}$}}
\Text(80,-10)[c]{\scalebox{0.7}[0.7]{$\Delta f_{1}\!-\!\Delta f_{2}$}}
\Text(30,-10)[c]{\scalebox{0.7}[0.7]{$\Delta f_{1}\!-\!\Delta f_{3}$}}
\includegraphics[width=8cm]{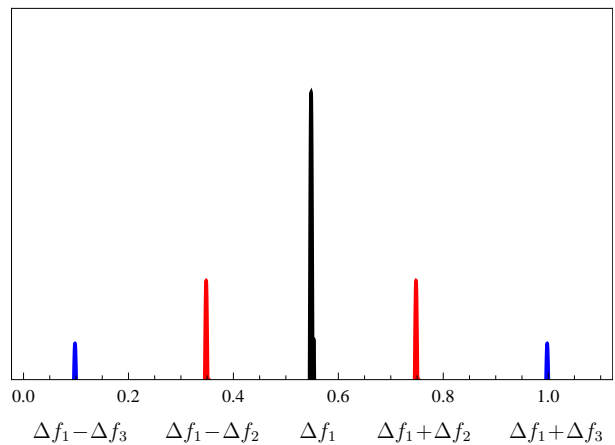}
\end{picture}
\end{center}
\vspace*{0.5cm}
\caption{Expected structure for a ``true'' signal (arbitrary units).
The black peak at the frequency $\Delta f_{1}$ is the signal. Our leakage monitoring
leads to sidebands at frequencies $\Delta f_{1}\pm\Delta f_{2}$ (red) and $\Delta f_{1}\pm \Delta f_{3}$ (blue) which are hopefully much smaller.}\label{Fig:signal}
\end{figure}

We can now turn to the detection. The first of our modulated signals with frequency $f_{0}+\Delta f_{1}$ (bottom of the picture) is
then fed into a mixer where it is combined, i.e., multiplied with the signal from the cavity (superheterodyne concept). This then passes through a low frequency band pass filter
is converted into a digital data with an analog digital converter (ADC) and is then recorded.
The last bandpass filter reduces the total amount of noise by eliminating the noise from high frequencies where we do not expect a signal.
The recorded signal will then be analyzed with the FFT method discussed in the previous section.
A proper signal should appear at frequency $\Delta f_{1}$ in the Fourier analysis (s. Figure~\ref{Fig:signal}).

%%%%%%%%%%%%%%%%%%%%%%%%%%%%%%%%%%%%%%%%%%%%%%%%%%%%%%%%
%%%%%%%%%%%%%%%%%%%%%%%%%%%%%%%%%%%%%%%%%%%%%%%%%%%%%%%%

Finally, let us discuss the leakage monitoring. In a set of frequency locked synthesizers we have created two additional frequencies $f_{0}+\Delta f_{2}$ and
$f_{0}+\Delta f_{3}$. We can use these for leakage monitoring.
Since we will look for unwanted leakages and corresponding resonances in our setup we will choose our test frequencies $f_{0}\pm\Delta f_{2,3}$ to
lie within the bandwidth of the resonant cavities.
The frequency $f_{0}+\Delta f_{3}$ can be directly connected to an antenna that emits this
signal outside the receiver box. Looking for them at a frequency $\Delta f_{1}\pm \Delta f_{3}$ in our signal spectrum (cf Fig.~\ref{Fig:signal})
we can compare this ``leaked'' signal to the
signal emitted from the antenna to determine the amount of shielding achieved (of course in the desirable case where we observe no signal at this
frequency we know the minimal amount of shielding). In a similar fashion we can  (optionally) put a small signal inside the outer box to monitor
the amount of shielding of the inner box. Taking the frequency of this test signal to be $f_{0}+\Delta f_{2}$ we would observe it
in the final spectrum at $\Delta f_{1}\pm \Delta f_{2}$ (see Fig.~\ref{Fig:signal}). We can also separately monitor the shielding provided by the outer box.
Inserting a receiver,
into the box (in Fig.~\ref{Fig:box1} upper right corner of the box) and putting the signal to a separate analyzer we can observe the leaked signals
at frequencies $\Delta f_{1}$ and $\Delta f_{1}\pm\Delta f_{3}$.

\begin{figure}[t]
\centerline{
\includegraphics[angle=-90,width=15cm]{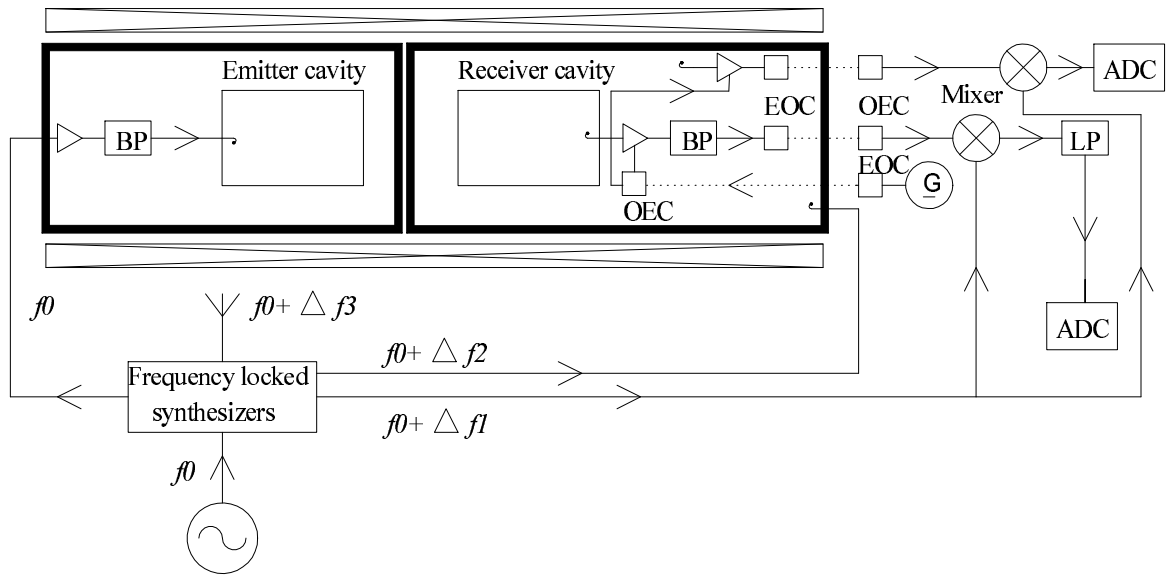} }
\caption{Setup for an experiment suitable for axion like particle detection. A solenoid provides for the required strong magnetic field.}\label{Fig:boxmagnet}
\end{figure}

So far we have looked at a setup to search for hidden photons. It is straightforward to modify the setup also to search for axion like particles.
The only difference is that here the conversion and reconversion of the photons requires a magnetic field. This can be provided by a suitable solenoid magnet.
This is shown in Fig.~\ref{Fig:boxmagnet}

%%%%%%%%%%%%%%%%%%%%%%%%%%%%%%%%%%%%%%%%%%%%%%%%%%%%%%%%
%%%%%%%%%%%%%%%%%%%%%%%%%%%%%%%%%%%%%%%%%%%%%%%%%%%%%%%%

\section{Example setups for HSPs and ALPs}

\begin{figure}[t]
\begin{center}
\subfigure{\includegraphics[width=7.5cm]{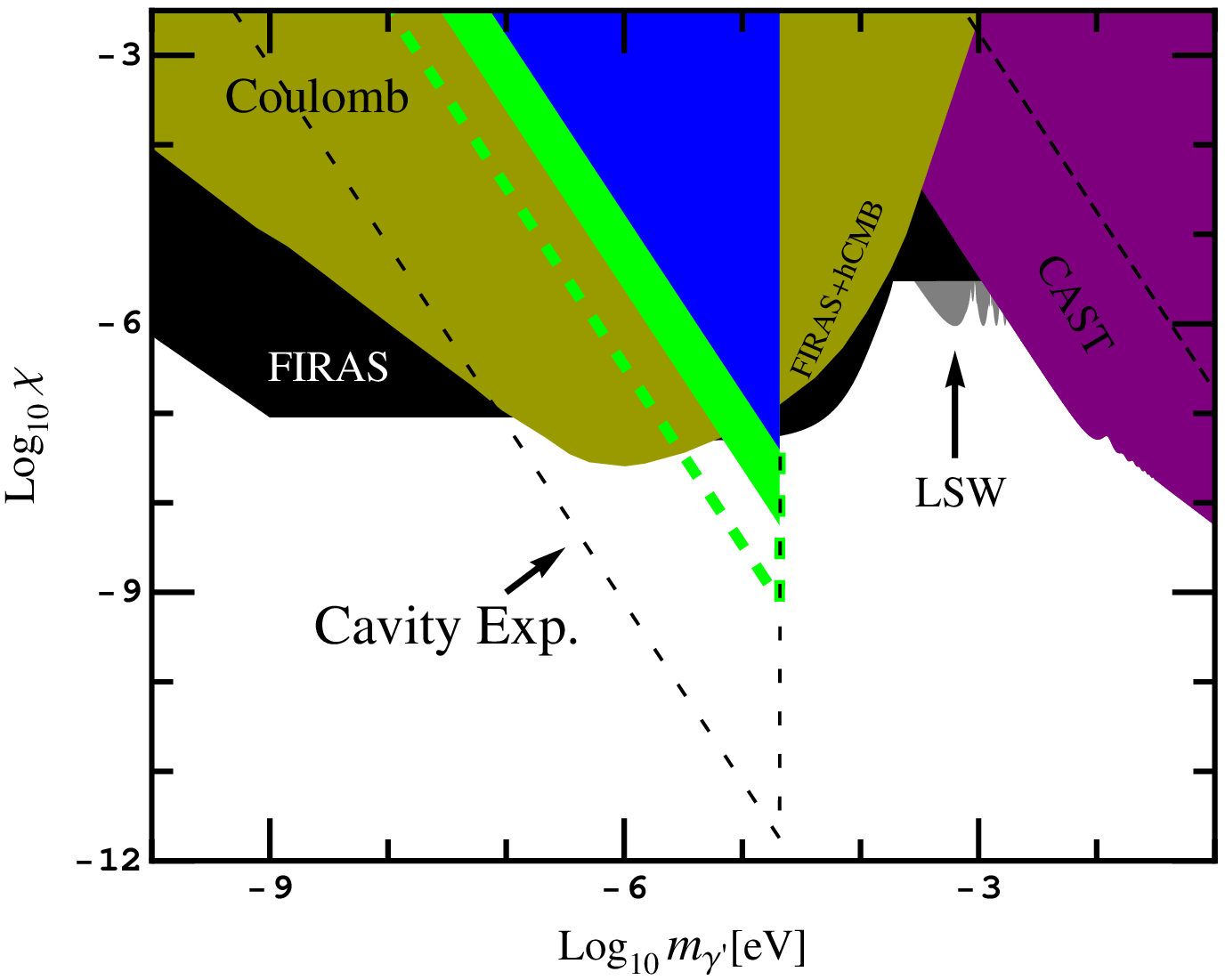} }
\hspace{0.5cm}
\subfigure{\includegraphics[width=7.5cm]{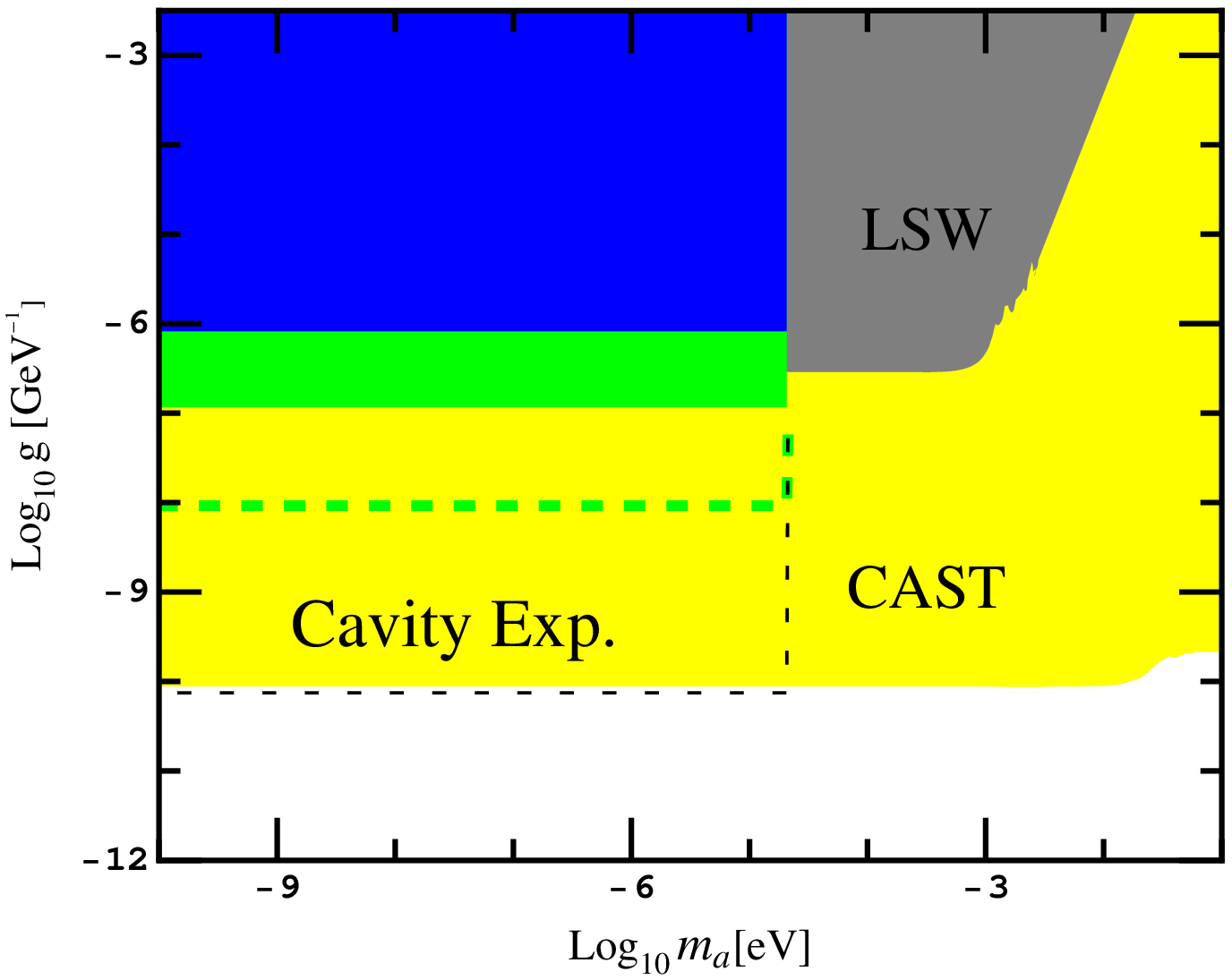} }
\end{center}
\caption{Sensitivity of the proposed experiment for hidden sector photons (left panel) and axion like particles (right panel).
The blue shaded areas correspond to a sensitivity of $P_{\rm det}=10^{-22}$~W, $Q_{1}=Q_{2}=2\times 10^3$ and
a power input of $P_{\rm in}=1000$~W. The green area is a setup with $P_{\rm det}=10^{-24}$~W, $Q_{1}=Q_{2}=10^4$.
The area above the green dashed line would be tested in a more advanced (but still realistic) setup with
$P_{\rm det}=10^{-26}$~W, $Q_{1}=Q_{2}=5\times 10^4$ and $P_{\rm in}=10^3$~W. We have used a frequency of 5 GHz a geometry factor $|G|=0.1$ and
for the axion curves a magnetic field of 5~T.
Finally, the thin dashed black lines correspond to
two more ambitious setups with $P_{\rm det}=10^{-28}$~W. For the hidden photon search we use superconducting cavities with,
$Q_{1}=Q_{2}=10^{10}$ and $P_{\rm in}=10$~W. For the axion search we stay with the normal conducting cavities $Q_{1}=Q_{2}=3\times 10^4$
but further increase the input power to $P_{\rm in}=10^5$~W and improve the magnetic field to 45~T and the geometry factor to $|G|=1$.
Also shown are current experimental limits on the possible existence of a hidden photon (left panel).
Strong constraints arise from the non-observation of deviations from
the Coulomb law (green-yellow)~\cite{Williams:1971ms,Bartlett:1988yy,Popov:1991,Popov:1999},
from Cosmic Microwave Background (CMB) measurements of the effective number of neutrinos and the blackbody nature
of the spectrum (black)~\cite{Jaeckel:2008fi,Mirizzi:2009iz}, from light-shining-through-walls (LSW)
experiments (grey)~\cite{Ruoso:1992nx,Cameron:1993mr,Robilliard:2007bq,Ahlers:2007rd,Chou:2007zzc,Ahlers:2007qf,Afanasev:2008jt,Fouche:2008jk,Afanasev:2008fv,Ehret:2009sq}, and from
searches of solar hidden photons with the CAST experiment (purple)~\cite{Andriamonje:2007ew,Redondo:2008aa}. The current limits on
axion like particles (right panel) arise from the CAST experiment~\cite{Arik:2008mq} and
LSW experiments~\cite{Ruoso:1992nx,Cameron:1993mr,Robilliard:2007bq,Chou:2007zzc,Afanasev:2008jt,Fouche:2008jk,Afanasev:2008fv,Ehret:2009sq}.}\label{Fig:hpsensitivity}
\end{figure}

Let us now see what sensitivities we can achieve using the technology discussed above and compare them to the current experimental bounds
shown in Fig.~\ref{Fig:hpsensitivity}.

%%%%%%%%%%%%%%%%%%%%%%%%%%%%%%%%%%%%%%%%%%%%%%%%%%%%%%%%
%%%%%%%%%%%%%%%%%%%%%%%%%%%%%%%%%%%%%%%%%%%%%%%%%%%%%%%%

As a first modest step we start with a setup at room temperature. Using the technology described above a detection sensitivity
of $10^{-22}$~W was reached with commercially available equipment in \cite{technical} with a measurement time of $300$~s and an average over 10 measurements.

The next important ingredient are the cavities, their quality and their frequency.
Since the maximal mass of a particle that can be tested in such an experiment is given by the frequency $m_{\rm max}=\omega_{0}=2\pi f_{0}$
we would like to choose the frequency as high as possible. For the hidden photon search high frequencies are also preferred because the current experimental bounds weaken
above masses of the order of $10^{-6}$~eV corresponding to frequencies above $240$~MHz.
The downside of high frequencies is that the $Q$ typically decreases as $Q\sim 1/\sqrt{f}$. As an example we choose $f=5$~GHz.
At this frequency a simple ``pillbox'' cavity in the lowest TM010 mode has a radius of $\sim 2$~cm and is therefore easy to handle and also fits inside a magnet.

The geometry factor will depend mainly on the chosen mode and the distance between the cavities. Typically the geometry factor is larger
for low modes such as the TM010 mode mentioned above (for a more in depth study see~\cite{tobar}).
For this mode the geometry factor for reasonable distances, say $d\lesssim 50$~cm is typically of the order of $G\sim 0.01-1$.
In the following we will simply assume $|G|=0.1$ for most setups.

To have optimal power input/output for the cavities we take cavities to be critically coupled to the generator and the receiver.
For the receiver this means that half the energy goes into the receiver and half is lost in the cavity. Therefore, the ``loaded'' $Q$ is exactly
half the ``empty'' $Q_{0}$ of the uncoupled cavity (similarly for the emitter cavity).
For the TM010 mode the quality factor is given by (neglecting surface roughness which typically decrease the quality factor by $20-30$\%),
\begin{equation}
\label{cavity}
2Q=Q_{0}=2.405\,c\sqrt{\frac{\sigma\mu_{0}}{4\pi f}}\frac{1}{1+x},
\end{equation}
where $\sigma$ is the conductivity of the cavity walls and $x$ is the ratio between radius and length of the cavity. For convenience we
have included a factor $c$ for the speed of light and a factor $\mu_{0}=1.26\times10^{-6}{\rm Vs/Am}$ for the permeability of vacuum,
since the conductivity is typically given in SI units.
Higher mode cavities often have a higher $Q$. Therefore, optimizing $|G|Q$ seems
like an interesting possibility to improve the sensitivity (cf.~\cite{tobar}). For simplicity we will continue with the TM010 mode.

At room temperature the conductivity of copper is $\sigma=5.8\times10^{7}\,{\rm S/m}$.
Inserting into Eq.~\eqref{cavity} we obtain $Q\sim 10^4$ at a frequency of 5 GHz.
Therefore, quality factors of the order of $Q=2\times 10^3$ are easy to achieve in practise for normal conducting copper cavities.
With these modest $Q$ values tuning the cavities to be in resonance is not problematic since the bandwidth of the cavity is of the order of MHz.

The achievable sensitivities for this setup is shown in Fig.~\ref{Fig:hpsensitivity} as the blue area. The left panel shows the sensitivity for
hidden sector photons. Note, that even with this very modest setup we would already probe a small region of previously unexplored parameter space.
To search for axion like particles we need in addition a magnetic field. Since we can rotate the cavity to obtain optimal overlap with the magnetic field we can choose
a more practicable solenoid field. A field strength of 5~T is straightforward with current technology (for example modern magnetic resonance imaging
systems for medical uses employ field strength in excess of 5~T).

Let us now turn to possible improvements. With a longer measurement time of the order of $10^4$~s it is realistic to
expect a sensitivity of the order of $10^{-24}$~W. Using high quality cavities with $Q\sim 10^4$ we can improve the sensitivity to the green region.
For hidden photons the experiment then already probes a significant part of unexplored parameter space. This could be further improved by measuring at various
different frequencies which would move the exclusion triangle horizontally (not shown). For the axion experiment this still quite modest setup already
significantly improves upon the current best purely laboratory limits from light-shining-through-walls experiments.

Going to low temperatures further significant improvements are possible. Due to the lower thermal noise a detection sensitivity of $10^{-26}$~W becomes realistic.
Also the quality of the cavities can be improved. At vanishing magnetic fields the conductivity can increase by a large factor called the residual resistive ratio $RRR$.
For ultra-pure copper values as high as $RRR\sim 10^4$ have been achieved~\cite{enss}. This would allow $Q$ to be of the order of $10^6$.
More realistic values are $RRR\sim 10^2$, corresponding to $Q\sim 10^5$.
Moreover, in the high magnetic fields required for the axion-like particle search
the conductivity of the material is reduced due to magnetoresistance. This can be roughly described by the K\"ohler law~\cite{Kohler,Caspers},
\begin{equation}
\sigma(B)=\frac{\sigma(0)}{1+0.002\left(\frac{B}{\rm T} RRR\right)^{1.055}}.
\end{equation}
At 5 Tesla and an $RRR=100$ the magnetoresistance reduces the possible $Q$ by about 30\%.
Using this $Q\sim 5\times 10^4$ seems plausible.
Finally, in order to not spent too much power on cooling we then have to reduce the power
input to $10^3$~W. With this setup we could test the parameter space above the green dashed line.

Finally, we can think about a more advanced setup. Moreover we now specialize for the specific particle species. For a hidden photon
search we can use superconducting cavities. These can have $Q\sim10^{10}$ at an input power of the order of 10~W~\cite{Aune:2000gb,Lilje:2004ib}.
Here, the precise tuning of the cavities
within their bandwidth $\sim 1$~Hz becomes more challenging. At a temperature of 2~K and measuring for about 5 days one might be able
to detect signals as low as $10^{-28}$~W. This would then test the area above the black dashed line, corresponding to an improvement of more than
four orders of magnitude beyond the current limits.
For the axion like particle search using superconducting cavities is problematic. However, we can improve the setup by using a stronger magnetic field.
Improving the magnetic field is especially effective since
the bound in the coupling is linear in $1/B$ (in contrast the bound scales as $(P_{\rm det})^{1/4}$).
Continuous fields as high as 45~T are possible \cite{emagnet}.
Due to magnetoresistance the $Q$ value decreases at these high field. Optimistically we can achieve $Q\sim 3\times10^4$. To compensate we
increase the power input to $10^5$~W (this is an extreme challenge for the cooling system).
Finally, we can also hope to use an improved geometry with $|G|=1$. With this (admittedly challenging) setup one can even venture beyond the current
best astrophysical bounds for axion like particles.

\section{Conclusions and outlook}
In this note we have argued that microwave cavity experiments can provide a powerful tool to search for weakly interacting sub-eV particles,
in particular for hidden sector photons and axion like particles.

Employing a narrowband detection method already with off the shelf tools from radio frequency technology detection sensitivities of the order of $10^{-23}$~W
have been demonstrated and $\lesssim 10^{-26}$ seem achievable. For typical frequencies in the GHz range this corresponds to sensitivities better than 1 photon per second.
To make full use of this sensitivity we have discussed a box-in-the-box setup that allows to achieve sufficient shielding and on-line monitoring for leaks during
the measurement.
Combining these ingredients with high quality cavities this setup allows to probe significant amounts of unexplored parameter space for hidden sector photons.
Adding a magnetic field we can also search for axion like particles and improve upon current laboratory bounds. With a more advanced setup
it also seems possible to improve upon the astrophysical bounds for axion like particles.

Further improvements could be achieved by improving the quality of the cavities. In particular for the axion like particle search one would like
to increase the quality of normal conducting cavities (the magnetic field cannot penetrate a superconductor). One tempting idea is to use cavities (partially)
filled with dielectric materials (cf. also~\cite{tobar}). This typically allows for much higher quality factors $Q\gtrsim 10^9$ even in normal conducting cavities.
However, typically the highest quality factors are achieved with relatively high order cavity modes which typically reduces the geometrical factor~\cite{tobar}.
Moreover, a dielectric medium has a tendency to decrease the production of hidden photons and axion like particles. Nevertheless, due to the enormous increase
in the quality factor a more careful investigation to optimize the setup with a dielectric medium seems promising.
An alternative route for improvements could be to use ``hard superconductors'' for the cavities. These type-II superconductors remain superconducting even when
penetrated by a magnetic field. If this can be realised it would open the way for improvements by several orders of magnitude in the sensitivity for axion like
particles.

Overall, already with relatively simple and cheap technology microwave cavity experiments can provide a powerful too
to search for new physics beyond the standard model in the form of weakly interacting sub-eV particles.

\section*{Acknowledgements}
The authors would like to thank K.~Zioutas for bringing the authors together. F.~C. would like to
thank R.~Heuer, S.~Myers and E.~Ciapala for encouragement and support.
J.~J. and A.~R. would like to thank A.~Gamp and A.~Lindner for stimulating discussions.

\end{document}